\documentclass{article}
\usepackage{spconf,amsmath,graphicx}

\usepackage{multirow}
\usepackage{float}
\usepackage{array}
\usepackage[table]{xcolor}
\newcolumntype{?}{!{\vrule width 1pt}}

\title{Domain Adaptation of low-resource Target-Domain models \newline using well-trained ASR Conformer Models}
\name{Vrunda N.Sukhadia, S. Umesh}
\address{Speech Lab, Dept. of Electrical Engineering, IIT Madras, Chennai, India}
\copyrightnotice{978-1-6654-7189-3/22/\$31.00~\copyright2023 IEEE}
\begin{document}
%
\maketitle
\begin{abstract}
In encoder-decoder framework for Automatic Speech Recognition (ASR) systems, the decoder of the well-trained ASR model is largely tuned towards the source-domain, hurting the performance of target-domain models in vanilla transfer-learning. On the other hand, the encoder layers of the well-trained ASR model mainly capture the acoustic characteristics. In this paper, the embeddings tapped from the encoder layers of a well-trained ASR model are used as features for domain adaptation of a downstream low resource Conformer target-domain model. We do ablation studies on optimal encoder layers for tapping embeddings and the effect of freezing or updating the well-trained ASR model’s encoder layers. Lastly, the application of Spectral Augmentation (SpecAug) on the proposed features improves the target-domain performance further. The proposed method reports an average relative improvement of $\sim$40\% over baseline with different source-domain model and target-domain Conformer model combinations.
\end{abstract}
\begin{keywords}
Domain adaptation, Automatic Speech Recognition (ASR), Pre-trained model, Low-resource speech recognition, 
Feature Extraction
\end{keywords}

\section{Introduction}

In most real-world problems, the amount of in-domain labeled data available for training Automatic Speech Recognition (ASR) models is usually very limited. Collecting a large amount of transcribed speech data is very expensive. Therefore, building in-domain ASR models with good accuracy is very difficult. One of the popular approaches to overcome this problem is to use transfer-learning, assuming a well-trained ASR model trained on a large dataset is available, possibly from a different domain. The transfer of knowledge can be in the form of feature representation, parameters, or weights \cite{pan2009survey}. 

Most state-of-the-art ASR models use encoder-decoder architecture. The decoder is optimized to learn the language model of the source-domain. In transfer-learning, when there is a mismatch in the domain, the performance of the transfer-learnt model would be affected. In \cite{meng2021internal,zeyer2021LibriSpeech,wang2020multitask}, the idea of internal language model estimation is used to improve performance when the well-trained ASR model is used to decode target-domain (i.e., in-domain) data directly without any transfer-learning step. While the decoder of the well-trained ASR model may be tuned toward the source-domain, the encoder layers may still model the acoustic aspects well, which can be transferred to the target-domain model. Therefore, in this paper, we investigate the use of encoder layer output of the well-trained ASR model as embeddings, which are fed to the randomly initialized encoder layers of the low-resource target-domain model instead of conventional Mel-filterbank features. Recently, pre-trained models such as Wav2Vec2.0 \cite{baevski2020wav2vec}, HuBERT \cite{hsu2021hubert} have become popular to address the low-resource speech modelling problem. However, they do suffer significant degradation when the target-domain is different from the data used in pre-training the model \cite{hsu2021robust,conneau2019unsupervised}. One approach is to use the target data along with the original pre-train data and start pre-training again. However, this is expensive. Loosely, the well-trained ASR models used in this paper are analogous to the pretrained self-supervised (SSL) models, except they have been trained in a supervised framework. The focus of this paper is, therefore, on domain adaptation to improve the performance of target-domain ASR models using a well-trained ASR model from a different domain. 

Our approach is along a line of thinking that bears some resemblance to the works in \cite{ghahremani2017investigation,yosinski2014transferable}.
Through a series of experiments, we investigate the effect of following different methods on the final performance:
\begin{itemize}
    \item The effect of different layer embeddings of the well-trained ASR model on the final performance of the target-domain model (i.e., in-domain model).
    \item The effect of freezing the well-trained ASR model's encoder layers versus allowing all the layers to update while training the low-resource target-domain model.
    \item The effect of applying a Spectral Augmentation (SpecAug) \cite{park2019specaugment} step on the well-trained ASR encoder features to improve the performance of the low-resource model. Note that well-trained ASR models have been trained after applying SpecAug on the Mel-filterbank features. The SpecAug is now applied when the ASR encoder embeddings are fed as features to the target-domain model.
\end{itemize}

The paper is organized as follows. Section \ref{sec:2} presents the details on the datasets and the architecture of source and target-domain models and establishes the baselines. Section \ref{sec:3} discusses the proposed method and shows the comparison with SSL features. Finally, the paper is concluded in Section \ref{sec:4}. 

\section{Datasets and Architecture Details}\label{sec:2}
\subsection{Datasets}
 In this paper, we consider several American English speech datasets from different domains. The datasets include:
\begin{itemize}
     \item \textit{SPGI} data consisting of narrated and spontaneous speech from the finance and earnings domain \cite{o2021spgispeech}
     \item \textit{LibriSpeech-960} and \textit{LibriSpeech-100-clean}, which are read audio books data. \cite{panayotov2015LibriSpeech}
     \item \textit{WSJ}, which is read Wall Street Journal news articles.
     \item \textit{GigaSpeech}, which is combination of audiobooks, podcasts and YouTube, covering both read and spontaneous speech. \cite{gigaspeech_data}
 \end{itemize}

\textit{LibriSpeech}-train-clean-100 [100 hours] is taken as the train data for \textit{LibriSpeech} as a target-data. \textit{LibriSpeech-dev-clean} [5.4 hours] is taken as the validation set and for testing \textit{LibriSpeech-dev-clean} [5.4 hours], \textit{LibriSpeech-dev-other} [5.3 hours], \textit{LibriSpeech-test-clean}  [5.4 hours] and \textit{LibriSpeech-test-other} [5.1 hours] are taken. \textit{WSJ-SI-284} [81 hours] is taken as the train data for \textit{WSJ}. \textit{WSJ-dev-93} [1 hour] is taken as the validation set and testing is done on both \textit{WSJ-eval-93} [0.41 hours] and \textit{WSJ-dev-93}. 
 All datasets are in 16kHz format.

The goal of this paper is to investigate improvement that can be obtained from domain adaptation ideas using well-trained ASR models. Specifically, we consider  different setups as follows:
\begin{itemize}
    \item \textit{Librispeech} (train-clean-100) and \textit{WSJ} (SI-284) datasets are the target-domain data.
    \item \textit{SPGI - 5000hrs}, \textit{LibriSpeech - 960hrs}, and \textit{GigaSpeech - 10000hrs} pre-trained ASR models are used as source-domain models. These models are readily available for download in ESPnet \cite{DBLP:journals/corr/abs-2012-13006}
\end{itemize}

In all scenarios, the source and target domains are different.

\subsection{Architecture Details}
All the ASR models built in this paper are based on the Conformer framework \cite{gulati2020conformer} with joint Connectionist Temporal Classification (CTC)/attention multi-task learning with a CTC weight of 0.3. All the pre-built conformer ASR models, namely, \textit{GigaSpeech-10000}, \textit{SPGI-5000} and \textit{LibriSpeech-960} models  are readily available on ESPnet GitHub page. The hyper-parameters used to train these pre-built ASR models are shown in Table~\ref{Pretrained model}. In the next subsection, we describe the baseline and vanilla transfer-learning using these pre-built models to obtain the \textit{WSJ} and \textit{Libri-100} target domain ASR model.

\begin{table}[h]
\begin{tabular}{c|c}
\hline \hline
\textbf{Hyperparameters} &\textbf{Values} \\ \hline \hline
Kernel Size                       &31     \\ \hline
Feature vector dimension          & 512  \\ \hline
Number of encoder layers          & 12   \\ \hline
Encoder units                     & 2048 \\ \hline
Number of decoder layers          & 6    \\ \hline
Decoder units                     & 2048 \\ \hline
Attention heads                   & 8    \\ \hline
Number of BPE(byte pair encoding) & 5000 \\ \hline
CTC weight                        & 0.3  \\ \hline 
\end{tabular}
\centering
\caption{Conformer Model Configuration for Well-trained ASR Models built using Large Labelled Datasets}
\label{Pretrained model}
\end{table}

\subsection{Baseline and Vanilla Transfer-Learning}
\subsubsection{Baseline}
The baseline Conformer models are built using the two low-resource target-domain datasets: \textit{LibriSpeech-100-clean} and \textit{WSJ SI-284} data.  We use 80-dimensional Mel-filterbank features with 3 extra pitch features which are extracted every 10 msec with 25 msec window length as input features for all experiments. We use 200 byte-pair encodings (BPE) in each case. The rest of the parameters are exactly the same as those used in the well-trained ASR models, as presented in Table \ref{Pretrained model}. All the layers are initialized with random initialization. No external Language Models are used while testing. Baseline results for \textit{LibriSpeech-100-clean} and \textit{WSJ SI-284} are presented in Tables \ref{LibriSpeech results} and \ref{wsj results} in Section 3, respectively. 
A layer-wise ablation study for the proposed approach facilitates the baseline results for different number of encoder layers to show that the improvement in the result is not because of increased number of encoder layers. A 22-6 encoder-decoder layers and a 24-6 encoder-decoder layers baseline is presented. We have also presented the results of directly decoding the target-domain data on the well-trained source-domain ASR model. The results have shown significant degradation compared to the baseline, indicating the mismatch between the source-domain of the well-trained ASR model and the target-domain data. 
One of the popular approaches to improve target-domain ASR performance is to do vanilla transfer-learning on the well-trained source-domain ASR model.

\subsubsection{Vanilla Transfer-learning}
We take the initialization of the target-domain ASR encoder-decoder model with the weights of the well-trained source-domain ASR model as vanilla transfer-learning. For vanilla transfer-learning, the 5000 BPEs of well-trained ASR models are unchanged. The encoder-decoder layers of the Conformer are then updated using the low-resource data. The results for vanilla transfer-learning are presented in Tables \ref{LibriSpeech results} and  \ref{wsj results} for both the scenarios under study. 

Though vanilla transfer-learning helps improve the performance, it might not be optimal for domain adaptation of an encoder-decoder model.
This is because the decoder of the well-trained source-domain model built on large datasets is optimized to learn the language model of the source-domain.

\section{Proposed Transfer-learning Approaches for Domain Adaptation}\label{sec:3}

\begin{figure}[h]
    \graphicspath{ {images/} }
    \includegraphics[width=7cm]{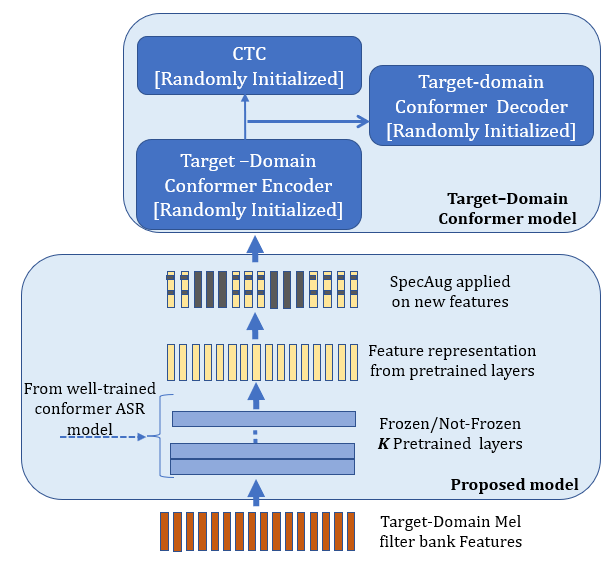}
    \centering
    \caption{Target-Domain Conformer ASR Model }
    \label{target-domain model}
    \centering
\end{figure}

We explore the use of embeddings tapped from encoder layers of well-trained ASR model instead of Mel-filterbank as input features for the target-domain Conformer ASR model. This is motivated by the fact that the encoder layers may model well the acoustic aspects, which may be transferred to the target-domain model. However, the decoder of the well-trained pre-built Conformer ASR model may be tuned towards the source-domain and may not be appropriate for domain adaptation. This problem of internal language model estimation of decoder has received quite a bit of attention over the past couple of years \cite{meng2021internal,zeyer2021LibriSpeech}. Therefore, to use the encoder-decoder of pre-trained ASR in the best way, we propose the following set-up:
\begin{itemize}
    \item The decoder in the source-domain well-trained model is heavily biased towards the source-domain text. It has learnt the internal language model for source-domain. As a result, if target domain data is mismatched, performance is degraded. 
    \item Training the decoder part using the target domain text data is one of the possible solutions. However, this is a challenging task for Attention-based Encoder-Decoder architecture as the decoder's training also depends on the cross-attention between the encoder-decoder. For this purpose, the decoder should get proper embedding from the encoder for respective text data.
    \item To address this problem, we add randomly initialised encoder layers on top of pre-trained ASR model encoder layers. To eliminate the influence of the biased decoder, we randomized the initialization of the decoder as well and jointly train entirely on the supervised target-domain data. This is shown in Fig.\ref{target-domain model}
    \item The above set-up of randomised encoder-decoder using pre-trained encoder embeddings is very similar to the approach employed for self-supervised models \cite{shinji,yang2021superb}. In their frameworks, embeddings extracted from the pre-trained Self-supervised model are passed through a conformer model to improve performance on the downstream task.
\end{itemize}

\begin{figure}[h]
    \graphicspath{ {images/} }
    \includegraphics[width=7cm]{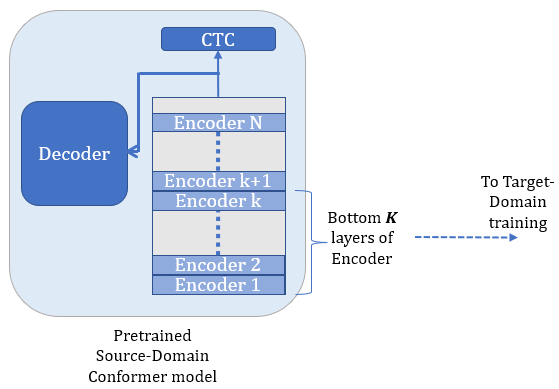}
    \centering
    \caption{Well-Trained Source-Domain Conformer ASR Model }
    \label{source-domain model}
    \centering
\end{figure}

Figure \ref{source-domain model} shows a well-trained source-domain Conformer ASR model. There are 12 encoder layers with the output at each layer being the 512-dimension embedding. The input features to the entire network are the Mel-filterbank features with SpecAug applied during training. The encoder layers capture more and more refined acoustic information, as we go up the encoder layers (of course, with some influence from the decoder layers). Conventional wisdom is that lower-layers capture more coarse acoustic units. Since these are well-trained models built on 10000-hour (or 5000/960-hour) data, the system has seen sufficient acoustic diversity. We plan to use outputs from these well-trained encoder layers as features (instead of Mel-filterbank features) to the target-domain Conformer model. This is shown in Fig. \ref{target-domain model}. 
A similar setup is often used for self-supervised models such as HuBERT and Wav2Vec2.0. However, it is well known that methods based on fine-tuning the self-supervised models often have degradation in performance when the target-domain data is different from pre-training data \cite{hsu2021robust}. This is because, self-supervised models are often trained on masked prediction criteria and therefore learn the sequence structure of acoustic units in pre-training data.

With our proposed set-up (shown in Fig.\ref{target-domain model}), we try to address two questions :
\begin{itemize}
    \item Which encoder layer output of Figure \ref{source-domain model} should be used in Figure \ref{target-domain model}, and how does it affect target-domain model performance?
    \item Should the well-trained encoder layers of pre-trained \textit{GigaSpeech} or \textit{SPGI-5000} or \textit{LibriSpeech-960}) be updated in Figure \ref{target-domain model} and how does it affect performance?
\end{itemize}
In the next two subsections, we address the above two questions.

\subsection{Effect of Layers from which Features are Tapped from Well-trained Conformer ASR Model}
We first test the efficacy of our approach where the bottom \textit{K} encoder layers are tapped and fed as features to the target-domain Conformer ASR model as shown in Figure \ref{target-domain model}. 
Tables \ref{LibriSpeech results} and \ref{wsj results} shows the effect of varying the tapping layer of well-trained Conformer ASR model on target-domain ASR performance.
As we tap from higher encoder layers, the ASR performance of target-domain ASR model improves for both the LibriSpeech target-domain model as well as the \textit{WSJ} model. All these results are without external Language Model to understand the acoustic modelling capability. 
As shown in Table \ref{LibriSpeech results}, relative improvement of $\sim$ 50\% and $\sim$30\% over the baseline is achieved  for \textit{clean} and \textit{other} datasets with \textit{GigaSpeech-10000} and \textit{SPGI} models, respectively, for \textit{LibriSpeech-100-clean} target-domain model. And as per the results shown in Table \ref{wsj results}, a relative improvement of $\sim$60\%, $\sim$50\%, and $\sim$50\% over baseline is achieved with \textit{GigaSpeech-10000}, \textit{SPGI}, and \textit{LibriSpeech-960} models, respectively for the \textit{WSJ} target-domain model. Moreover, there is a significant improvement over vanilla transfer-learning.
\subsection{Freezing vs. Updating the Encoder Layers from which Features are Tapped} \label{sec: 3.2}
Next, we allow all the layers to update 
during the training of target-domain ASR model and compare it with the results of freezing the lower encoder layers discussed in the previous section. The results are presented in Tables \ref{LibriSpeech results} and \ref{wsj results}.

\begin{table}[h]
\centering
\footnotesize
\setlength\tabcolsep{3.5pt}
\begin{tabular}{p{0.15\columnwidth}p{0.1\columnwidth}p{0.1\columnwidth}p{0.1\columnwidth}p{0.1\columnwidth}p{0.1\columnwidth}p{0.1\columnwidth}p{0.1\columnwidth}p{0.05\columnwidth}}
\hline\hline
\multicolumn{1}{c|}{\cellcolor{gray!30}\textbf{Approach}} &
  \multicolumn{2}{c|}{\cellcolor{gray!30}\textbf{\begin{tabular}[c]{@{}c@{}}test-\\clean\end{tabular}}} &
  \multicolumn{2}{c|}{\cellcolor{gray!30}\textbf{\begin{tabular}[c]{@{}c@{}}dev-\\clean\end{tabular}}} &
  \multicolumn{2}{c|}{\cellcolor{gray!30}\textbf{\begin{tabular}[c]{@{}c@{}}test-\\other\end{tabular}}} &
  \multicolumn{2}{c}{\cellcolor{gray!30}\textbf{\begin{tabular}[c]{@{}c@{}}dev-\\other\end{tabular}}} \\ 
 \hline \hline
\multicolumn{1}{c|}{Baseline} &
  \multicolumn{2}{c|}{7.8} &
  \multicolumn{2}{c|}{7.3} &
  \multicolumn{2}{c|}{21.7} &
  \multicolumn{2}{c}{21} \\ \hline
\multicolumn{1}{c|}{\begin{tabular}[c]{@{}c@{}}22 Enc layers \\ Baseline\end{tabular}} &
  \multicolumn{2}{c|}{7.7} &
  \multicolumn{2}{c|}{7.3} &
  \multicolumn{2}{c|}{20.9} &
  \multicolumn{2}{c}{20.2} \\ \hline
\multicolumn{1}{c|}{\begin{tabular}[c]{@{}c@{}}24 Enc layers\\  Baseline\end{tabular}} &
  \multicolumn{2}{c|}{8} &
  \multicolumn{2}{c|}{7.6} &
  \multicolumn{2}{c|}{21.3} &
  \multicolumn{2}{c}{20.8} \\ \hline \hline
  \multicolumn{1}{c?}{\cellcolor{gray!30}} &
  \multicolumn{4}{c?}{\cellcolor{gray!30}\textbf{GigaSpeech}} &
  \multicolumn{4}{c}{\cellcolor{gray!30}\textbf{SPGI}} \\\cline{2-9}  
  \multicolumn{1}{c?}{\multirow{-2}{*}{\cellcolor{gray!30}\textbf{Approach}}}&
  \multicolumn{1}{c|}{\cellcolor{gray!30}\begin{tabular}[c]{@{}c@{}}test-\\clean\end{tabular}} &
  \multicolumn{1}{c|}{\cellcolor{gray!30}\begin{tabular}[c]{@{}c@{}}dev-\\clean\end{tabular}} &
  \multicolumn{1}{c|}{\cellcolor{gray!30}\begin{tabular}[c]{@{}c@{}}test-\\other\end{tabular}} &
  \multicolumn{1}{c?}{\cellcolor{gray!30}\begin{tabular}[c]{@{}c@{}}dev-\\other\end{tabular}} &
\multicolumn{1}{c|}{\cellcolor{gray!30}\begin{tabular}[c]{@{}c@{}}test-\\clean\end{tabular}} &
  \multicolumn{1}{c|}{\cellcolor{gray!30}\begin{tabular}[c]{@{}c@{}}dev-\\clean\end{tabular}} &
  \multicolumn{1}{c|}{\cellcolor{gray!30}\begin{tabular}[c]{@{}c@{}}test-\\other\end{tabular}} &
  \multicolumn{1}{c}{\cellcolor{gray!30}\begin{tabular}[c]{@{}c@{}}dev-\\other\end{tabular}} \\ \hline \hline
\multicolumn{1}{c?}{\begin{tabular}[c]{@{}c@{}}Direct \\ Decoding\end{tabular}} &
  \multicolumn{1}{c|}{18.2} &
  \multicolumn{1}{c|}{15.3} &
  \multicolumn{1}{c|}{21.2} &
  \multicolumn{1}{c?}{20.9} &
  \multicolumn{1}{c|}{23.5} &
  \multicolumn{1}{c|}{23.1} &
  \multicolumn{1}{c|}{31.6} &
  31.1 \\ \hline
\multicolumn{1}{c?}{\begin{tabular}[c]{@{}c@{}}Vanilla\\ Transfer-\\ learning\end{tabular}} &
  \multicolumn{1}{c|}{6.6} &
  \multicolumn{1}{c|}{5.7} &
  \multicolumn{1}{c|}{11.5} &
  \multicolumn{1}{c?}{11.9} &
  \multicolumn{1}{c|}{7.6} &
  \multicolumn{1}{c|}{7.4} &
  \multicolumn{1}{c|}{21.2} &
  20.9 \\ \hline
\multicolumn{9}{c}{\cellcolor{gray!15}Updating Layers} \\ \hline
\multicolumn{1}{c?}{6 Layers} &
  \multicolumn{1}{c|}{5.9} &
  \multicolumn{1}{c|}{5.4} &
  \multicolumn{1}{c|}{15.8} &
  \multicolumn{1}{c?}{16.0} &
  \multicolumn{1}{c|}{6.7} &
  \multicolumn{1}{c|}{6.4} &
  \multicolumn{1}{c|}{18} &
  17.9 \\ \hline
\multicolumn{1}{c?}{8 Layers} &
  \multicolumn{1}{c|}{5.8} &
  \multicolumn{1}{c|}{5.6} &
  \multicolumn{1}{c|}{15.7} &
  \multicolumn{1}{c?}{16.0} &
  \multicolumn{1}{c|}{6.5} &
  \multicolumn{1}{c|}{6.3} &
  \multicolumn{1}{c|}{17.4} &
  17.3 \\ \hline
\multicolumn{1}{c?}{10 Layers} &
  \multicolumn{1}{c|}{5.2} &
  \multicolumn{1}{c|}{4.8} &
  \multicolumn{1}{c|}{13.9} &
  \multicolumn{1}{c?}{14.3} &
  \multicolumn{1}{c|}{5.9} &
  \multicolumn{1}{c|}{5.7} &
  \multicolumn{1}{c|}{16} &
  15.8 \\ \hline
\multicolumn{1}{c?}{12 Layers} &
  \multicolumn{1}{c|}{\textbf{3.9}} &
  \multicolumn{1}{c|}{\textbf{3.8}} &
  \multicolumn{1}{c|}{\textbf{10.5}} &
  \multicolumn{1}{c?}{\textbf{10.9}} &
  \multicolumn{1}{c|}{\textbf{5.5}} &
  \multicolumn{1}{c|}{\textbf{5.3}} &
  \multicolumn{1}{c|}{14.8} &
  14.9 \\ \hline
\multicolumn{9}{c}{\cellcolor{gray!15}Freezing Layers} \\ \hline
\multicolumn{1}{c?}{6 Layers} &
  \multicolumn{1}{c|}{6.0} &
  \multicolumn{1}{c|}{5.8} &
  \multicolumn{1}{c|}{14.1} &
  \multicolumn{1}{c?}{13.9} &
  \multicolumn{1}{c|}{6.8} &
  \multicolumn{1}{c|}{6.5} &
  \multicolumn{1}{c|}{16.8} &
  16.4 \\ \hline
\multicolumn{1}{c?}{8 Layers} &
  \multicolumn{1}{c|}{5.4} &
  \multicolumn{1}{c|}{5.2} &
  \multicolumn{1}{c|}{12.4} &
  \multicolumn{1}{c?}{12.5} &
  \multicolumn{1}{c|}{6.9} &
  \multicolumn{1}{c|}{6.4} &
  \multicolumn{1}{c|}{15.3} &
  14.8 \\ \hline
\multicolumn{1}{c?}{10 Layers} &
  \multicolumn{1}{c|}{4.7} &
  \multicolumn{1}{c|}{4.5} &
  \multicolumn{1}{c|}{10.9} &
  \multicolumn{1}{c?}{11.1} &
  \multicolumn{1}{c|}{6.7} &
  \multicolumn{1}{c|}{6.4} &
  \multicolumn{1}{c|}{14.7} &
  14.2 \\ \hline
\multicolumn{1}{c?}{12 Layers} &
  \multicolumn{1}{c|}{4.7} &
  \multicolumn{1}{c|}{4.3} &
  \multicolumn{1}{c|}{11.0} &
  \multicolumn{1}{c?}{11.0} &
  \multicolumn{1}{c|}{6.5} &
  \multicolumn{1}{c|}{6.4} &
  \multicolumn{1}{c|}{\textbf{14.5}} &
  \textbf{14.2} \\ \hline
\end{tabular}
\caption{Comparison of Freezing and Updating the Feature Extractor as  Encoder-Layer \textit{K} varies for \textit{LibriSpeech-100-clean} Target Model with Encoder Layers from \textit{GigaSpeech-10000} and \textit{SPGI-5000}. No External Language Model was used.}
\label{LibriSpeech results}
\end{table}

\begin{table}[h]
\centering
\footnotesize
\setlength\tabcolsep{7pt}
\begin{tabular}{ccccccc}
\hline \hline
  \multirow{6}{*}{\textbf{Approach}} 
   &
  \multicolumn{6}{|c}{\textbf{Model}} \\ \cline{2-7}
   &
  \multicolumn{2}{?c}{\textbf{GigaSpeech}} &
  \multicolumn{2}{?c}{\textbf{SPGI}} &
  \multicolumn{2}{?c}{\textbf{\begin{tabular}[c]{@{}c@{}}LibriSpeech-\\ 960\end{tabular}}} \\ 
  \cline{2-7}
  &
  \multicolumn{1}{?c|}{\begin{tabular}[c]{@{}c@{}}test-\\ dev\\ 93\end{tabular}} &
  \multicolumn{1}{c}{\begin{tabular}[c]{@{}c@{}}test-\\ eval\\ 93\end{tabular}} &
  \multicolumn{1}{?c|}{\begin{tabular}[c]{@{}c@{}}test-\\ dev\\ 93\end{tabular}} &
  \multicolumn{1}{c}{\begin{tabular}[c]{@{}c@{}}test-\\ eval\\ 93\end{tabular}} &
  \multicolumn{1}{?c|}{\begin{tabular}[c]{@{}c@{}}test-\\ dev\\ 93\end{tabular}} &
  \begin{tabular}[c]{@{}c@{}}test-\\ eval\\ 93\end{tabular} \\
  \hline \hline
\multicolumn{1}{c}{Baseline} &
  \multicolumn{1}{?c|}{13.6} &
  \multicolumn{1}{c}{11.9} &
  \multicolumn{1}{?c|}{13.6} &
  \multicolumn{1}{c}{11.9} &
  \multicolumn{1}{?c|}{13.6} &
  \multicolumn{1}{c}{11.9}\\ \hline
\multicolumn{1}{c}{\begin{tabular}[c]{@{}c@{}}Direct \\ Decoding\end{tabular}} &
  \multicolumn{1}{?c|}{13.2} &
  \multicolumn{1}{c}{12.4} &
  \multicolumn{1}{?c|}{23.4} &
  \multicolumn{1}{c}{23.3} &
  \multicolumn{1}{?c|}{30.9} &
   30.3\\ \hline
\multicolumn{1}{c}{\begin{tabular}[c]{@{}c@{}}Vanilla\\ Transfer-\\ learning\end{tabular}} &
  \multicolumn{1}{?c|}{7.8} &
  \multicolumn{1}{c}{6.4} &
  \multicolumn{1}{?c|}{13.5} &
  \multicolumn{1}{c}{7.4} &
  \multicolumn{1}{?c|}{8.9} &
  \multicolumn{1}{c}{7.2} \\ \hline
\multicolumn{7}{c}{Updating Layers} \\ \hline
\multicolumn{1}{c}{6 Layers} &
  \multicolumn{1}{?c|}{9.7} &
  \multicolumn{1}{c}{8.9} &
  \multicolumn{1}{?c|}{10.3} &
  \multicolumn{1}{c}{8.9} &
  \multicolumn{1}{?c|}{10.4} &
  \multicolumn{1}{c}{8.5} \\ \hline
\multicolumn{1}{c}{8 Layers} &
  \multicolumn{1}{?c|}{8.3} &
  \multicolumn{1}{c}{6.9} &
  \multicolumn{1}{?c|}{9.4} &
  \multicolumn{1}{c}{8.3} &
  \multicolumn{1}{?c|}{9.3} &
  \multicolumn{1}{c}{8.1} \\ \hline
\multicolumn{1}{c}{10 Layers} &
  \multicolumn{1}{?c|}{6.4} &
  \multicolumn{1}{c}{\textbf{4.4}} &
  \multicolumn{1}{?c|}{8.4} &
  \multicolumn{1}{c}{6.7} &
  \multicolumn{1}{?c|}{7.6} &
  \multicolumn{1}{c}{6.1} \\ \hline
\multicolumn{1}{c}{12 Layers} &
  \multicolumn{1}{?c|}{\textbf{6.2}} &
  \multicolumn{1}{c}{4.5} &
  \multicolumn{1}{?c|}{\textbf{7.4}} &
  \multicolumn{1}{c}{\textbf{5.6}} &
  \multicolumn{1}{?c|}{\textbf{6.8}} &
  \multicolumn{1}{c}{\textbf{5.5}} \\ \hline
\end{tabular}
\caption{Ablation study of Updating the Feature Extractor as  Encoder-Layer \textit{K} varies for \textit{WSJ-SI-284} Target Model with Encoder Layers from \textit{GigaSpeech-10000}, \textit{SPGI-5000} and, \textit{LibriSpeech-960}. No External Language Model was Used. }
\label{wsj results}
\end{table}

In this ablation study for \textit{LibriSpeech}, the experiments are performed with both updating and freezing the encoder layers. We observed that updating layers provide the best performance for the matched case test data. \textit{WSJ} test dataset, which has mostly read modern business English, has matched case test data. Hence, for \textit{WSJ} as the target-domain, we have shown only the ablation study for updating encoder layers of the pre-trained models.

From the Tables we observe that,

\begin{itemize}

    \item In both the scenarios, as we tap from higher and higher encoder layers, the target-domain ASR performance improves irrespective of freezing or updating encoder layers. 
    \item  For lower-layer tapping, say from $4$ or $6$, although the ASR performance is still better than the baseline, it degrades compared to vanilla transfer-learning. This is because the lower encoder-layers capture more coarse acoustics and the more high-level information has to be captured using only the 100-hour target-domain data.  

    \item When the encoder layers of the well-trained ASR Conformer model are allowed to update, matched datasets give better performance. This is evident from the results as the \textit{clean} sets like \textit{dev-clean} and \textit{test-clean} show improvement when the encoder layers are allowed to update over freezing, when trained with the \textit{clean} set \textit{LibriSpeech-100-clean}.
    \item For \textit{LibriSpeech-dev-other} and \textit{LibriSpeech-test-other}, which are data from lower quality recordings and have some non-US English accents, freezing the encoder layers of well-trained ASR model gives better ASR performance compared to updating the encoder layers.
\end{itemize}
From the mentioned observations, we can infer that 
\begin{itemize}
    \item Updating all the encoder layers of well-trained ASR model using target-domain data, the model becomes more tuned towards the target-domain. In such cases, it works very well for matched target test conditions.
  \item Freezing the encoder layers of the well-trained ASR model retains generalization capability since the well-trained model is trained on large data with sufficient diversity and helps mismatched test conditions such as \textit{dev-other} and \textit{test-other}.
      

\end{itemize}

\subsection{Comparison of Proposed Features of Supervised Model with the Features of SSL Models}
In these experiments, the proposed method is compared with the results of SSL method. For the comparison, features from SSL models are taken to train the target-domain Conformer encoder-decoder model. For the fair comparison, following readily available pre-trained models are taken which have comparable parameter count:
\begin{itemize}
    \item Supervised model trained on 960 hr \textit{LibriSpeech} data [Parameters: \textbf{78M}]
    \item \textit{Wav2Vec2.0} Base model trained on 960hr \textit{LibriSpeech} data [Parameters: \textbf{95M}]
    \item \textit{HuBERT} Base model trained on 960hr \textit{LibriSpeech} data [Parameters: \textbf{95M}]
    \item \textit{WavLM} Base model trained on 960hr \textit{LibriSpeech} data [Parameters: \textbf{95M}]
\end{itemize}

Features are extracted from these models and fed to the Conformer model which will be trained and tested on the \textit{WSJ} data. The target-domain Conformer model has same configuration as used in all the above experiments. WER for the target-domain model using different pre-trained model is shown in Table \ref{comparison with ssl}. The approach "10 Layers" and "12 Layers" in Table \ref{comparison with ssl} indicate that the features are extracted from the 10$^\text{th}$ layer and 12$^\text{th}$ layer of the supervised \textit{LibriSpeech 960hr} model, respectively. As shown in the results, "10 Layers" (Supervised) features improves $\sim$14\%, $\sim$15\% and $\sim$8\% WER over the \textit{Wav2Vec2.0}, \textit{HuBERT} and \textit{WavLM} features respectively. And the "12 Layers" (Supervised) features improve $\sim$22\% $\sim$23\% and $\sim$17\% WER over the \textit{Wav2Vec2.0}, \textit{HuBERT} and \textit{WavLM} features respectively. From these experiments it can be inferred that self supervised features perform better for in-domain target data but it suffers when it is exposed to out of domain target data. On the other hand, features extracted from the supervised model are better than self-supervised features in terms of domain shift. These results concur with the idea presented in \cite{hsu2021robust}.

\begin{table}[h]
\begin{tabular}{c|cc}
\hline \hline
\multirow{2}{*}{\textbf{Approach}} & \multicolumn{2}{c}{\textbf{WER}}                       \\ \cline{2-3} 
                          & \multicolumn{1}{c|}{test-dev93} & test-eval93 \\ \hline \hline

Base \textit{Wav2Vec2.0}           & \multicolumn{1}{c|}{8.7}          & 7.1           \\ \hline
Base \textit{HuBERT}               & \multicolumn{1}{c|}{8.7}          & 7.2           \\ \hline
Base \textit{WavLM}                & \multicolumn{1}{c|}{7.8}          & 6.7          \\ \hline
10 Layers                 & \multicolumn{1}{c|}{7.6}          & 6.1           \\ \hline
12 Layers                 & \multicolumn{1}{c|}{\textbf{6.8}}          & \textbf{5.5}      \\
\hline    
\end{tabular}
\centering
\caption{Comparison of Proposed Method with SSL Models for Domain Adaptation, with \textit{LibriSpeech-960hr} source-domain and \textit{WSJ-SI-284} target-domain model. No External Language Model was used.}
\label{comparison with ssl}
\end{table}

\subsection{Applying SpecAug on top of Encoder Layers of Well-trained Conformer ASR Model}

We explore the effect of applying SpecAug on top of the embeddings tapped from the encoder layers of the well-trained ASR model, before it is applied as features to the target-domain ASR model. It is to be noted that a SpecAug is already applied on the input Mel-filterbank features by default while training the ASR models. 
A similar idea is also used in the downstream tasks of self-supervised models such as \textit{Wav2Vec2.0} or \textit{HuBERT}. During the downstream ASR task, a type of SpecAug is present immediately after the Convolutional Neural Network (CNN) feature encoder of the self-supervised model.  

Since we observed that freezing the encoder layers gives better generalization on \textit{noisy} data, we have frozen the encoder layers for further experiments. Since SpecAug masks can be applied on Time and Frequency axis, the ablation study for the different mask width on Time and Frequency axis is shown in the Table \ref{SpecAug results}. From Table \ref{SpecAug results} we observe that a SpecAug after the encoder layers of well-trained ASR model, in addition to the SpecAug on Mel-filterbank features, gives better results compared to the results in Subsection \ref{sec: 3.2}. We have not conducted the experiment for 12 layers as there will be no scope for the target-domain encoder to recover from SpecAug distortions. But it can be seen from Table \ref{SpecAug results} that the performance of 10 layers with additional SpecAug is better than that of the 12 layers results in Subsection \ref{sec: 3.2}.


\begin{table}[h]
\begin{tabular}{c|c|c|c|c|c}
\hline \hline
\multicolumn{2}{c|}{\textbf{\begin{tabular}[c]{@{}c@{}}SpecAug\\ Mask-Width\end{tabular}}} &
  \multirow{2}{*}{\textbf{\begin{tabular}[c]{@{}c@{}}Test\\ clean\\\textbf{WER}\\{ \%}\end{tabular}}} &
  \multirow{2}{*}{\textbf{\begin{tabular}[c]{@{}c@{}}Dev\\ clean\\\textbf{WER}\\ \%\end{tabular}}} &
  \multirow{2}{*}{\textbf{\begin{tabular}[c]{@{}c@{}}Test\\ other\\\textbf{WER}\\ \%\end{tabular}}} &

  \multirow{2}{*}{\textbf{\begin{tabular}[c]{@{}c@{}}Dev\\ other\\\textbf{WER}\\ \%\end{tabular}}} \\ \cline{1-2}
\textbf{\begin{tabular}[c]{@{}c@{}}Frequency \\Mask\\  Width\end{tabular}} &
  \textbf{\begin{tabular}[c]{@{}c@{}}Time \\ Mask \\ Width\end{tabular}} &
  &
  &
  &
  \\ \hline \hline
\multicolumn{6}{c}{6 pre-trained layers}                           \\ \hline
\multicolumn{2}{c|}{Without SpecAug} & 6.8  & 6.5 & 16.8    & 16.4  \\ \hline
20                & 10                & 6.4  & \textbf{6.2} & 15.5    & 15.0  \\ \hline
20                & 20                & \textbf{6.4}  & 6.3 & \textbf{15.1}    & \textbf{14.9}  \\ \hline
\multicolumn{6}{c}{10 pre-trained layers}                          \\ \hline
\multicolumn{2}{c|}{Without SpecAug} & 6.7  & 6.4 & 14.7    & 14.2  \\ \hline
20                & 10                & 6.2  & 6.1 & 13.7    & 13.4  \\ \hline
20                & 20                & \textbf{6.2}   & \textbf{6.0} & \textbf{13.6}    & \textbf{13.1}  \\ \hline
\end{tabular}
\centering
\caption{Ablation Study for SpecAug Mask Width for \textit{LibriSpeech-100-clean}}
\label{SpecAug results}
\end{table}

\section{Conclusion}\label{sec:4}
In this paper we have investigated methods to optimally utilize the potential of an ASR model trained with a large dataset in training a new ASR model with a low-resource data. Though vanilla transfer-learning gives better performance than the baseline, the results of our proposed transfer-learning approaches show that vanilla transfer-learning does not tap the complete potential of well-trained ASR model, when the target data is from a new domain.  This is due to the strong Language Model learnt by the decoder of the well-trained ASR model. Our approach is to tap the features from the bottom \textit{K} layers of encoder of the well-trained ASR model and feed it to the low-resource ASR model. Our approach has shown significant improvements over the baseline and vanilla transfer-learning as \textit{K} increases. The results also indicate that freezing the encoder layers of the well-trained ASR model makes the target-domain model more robust. When those layers are updated, performance is better than freezing, if both the train and test datasets are matched. It can also be concluded from the experiments that the proposed approach is more robust to domain shift compared to the self-supervised model features. An additional SpecAug applied on the well-trained model's encoder features helps to further improve the ASR performance.

\bibliographystyle{IEEEtran}
\bibliography{ref}



\end{document}